\documentclass[twocolumn,english,aps,prb,floatfix]{revtex4}
\usepackage[T1]{fontenc}
\usepackage[latin2]{inputenc}
\usepackage{graphicx}
\usepackage{amssymb}

\makeatletter

\usepackage{bm}

\usepackage{babel}
\makeatother
\begin{document}

\title{Conductance of nanosystems with interaction}

\author{A. Ram\v{s}ak$^{1,2}$ and T. Rejec$^{1}$}

\affiliation{$^{1}$Jo\v{z}ef Stefan Institute, Ljubljana, Slovenia\\
 $^{2}$Faculty of Mathematics and Physics, University of Ljubljana,
Ljubljana, Slovenia }

\begin{abstract}
The zero-temperature linear response conductance through an interacting
mesoscopic region attached to noninteracting leads is investigated.
We present a set of formulas expressing the conductance in terms of
the ground-state energy of an auxiliary system, namely a ring threaded
by a magnetic flux and containing the correlated electron region.
We prove that the formalism is exact if the ground state of the system
is a Fermi liquid. We show that in such systems the ground-state energy
is a universal function of the magnetic flux, where the conductance
is the relevant parameter. The method is illustrated with results
for the transport through an interacting quantum dot and a simple
Aharonov-Bohm ring with Kondo-Fano resonance physics.
\end{abstract}
\maketitle

\section{introduction}

In the last decade technological advances enabled controlled fabrication
of small regions connected to leads and \textit{\emph{the conductance}}\textit{,}
\textit{\emph{relating}} the current through such a system to the
voltage applied between the leads, proved to be the most important
property of such systems. There is a number of such examples, e.g.
metallic islands prepared by e-beam lithography or small metallic
grains,\cite{Delft01} semiconductor quantum dots,\cite{Kouwenhoven97}
or a single large molecule such as a carbon nanotube or DNA. It is
possible to break a metallic contact and measure the transport properties
of an atomic-size bridge that forms in the break,\cite{Agrait02}
or even measure the conductance of a single hydrogen molecule.~\cite{Smit02}
Recent measurements of conductance through single molecules proved
that strong electron correlations can play an important role in such
systems.~\cite{nature}

The transport in noninteracting mesoscopic systems is theoretically
well described in the framework of the Landauer-B\"{u}ttiker formalism.
The conductance $G$ is at zero temperature determined with the Landauer-B\"{u}ttiker
formula~\cite{Landauer57} 

\begin{eqnarray}
 &  & G=G_{0}\left|t\left(\epsilon_{F}\right)\right|^{2},\quad\quad G_{0}=\frac{2e^{2}}{h}.\label{eq:LB}\end{eqnarray}
The key quantity here is the single particle transmission amplitude
$t(\epsilon_{F})$ for electrons at the Fermi energy. The formula
proved to be very useful and reliable, as long as electron-electron
interaction in a sample is negligible. However, the Landauer-B\"{u}ttiker
formalism cannot be directly applied to systems where the interaction
between electrons plays an important role. Several approaches have
been developed to allow one to treat such systems. The Kubo formalism
provides us with a conductance formula which is applicable in the
linear response regime and was intensively studied by Oguri.\cite{Oguri97a,Oguri01}
A more general approach applicable also to non-equilibrium cases was
developed by Meir and Wingreen.~\cite{Meir92} Recently, ab initio
methods to study the transport through small molecular junctions were
also applied.\cite{abinitio}

\section{Conductance formulas for Fermi liquid systems}

\begin{figure}[htbp]
\begin{center}\includegraphics[%
  width=6cm,
  keepaspectratio]{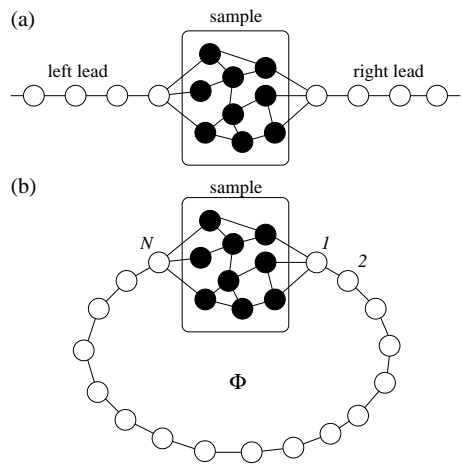}\end{center}

\caption{\label{cap:System1}(a) Schematic picture of a sample with interaction
connected to noninteracting leads. (b) The sample embedded in a ring
formed by joining the left and right leads of the system (a). Auxiliary
magnetic flux $\Phi=\frac{\hbar}{e}\phi$ penetrates the ring. }
\end{figure}

The relevant system is schematically presented in Fig.~\ref{cap:System1}(a).
A mesoscopic interacting region, which could be a molecule, a quantum
dot, a quantum dot array or a similar 'artificial molecule' system,
is attached to noninteracting leads. As shown in Ref.~\onlinecite{RR}
(hereafter referred to as RR), the conductance of such a system can
be determined solely from the ground-state energy of an auxiliary
system, formed by connecting the leads of the original system into
a ring and threaded by a magnetic flux, Fig.~\ref{cap:System1}(b).
The main advantage of this method is the fact that it is often much
easier to calculate the ground-state energy (for example, using variational
or quantum Monte Carlo methods) than the Green's function, which is
needed in the Kubo and Keldysh approaches. The method is applicable
only to a certain class of systems, namely to those exhibiting Fermi
liquid properties, at zero temperature and in the linear response
regime. However, in this quite restrictive domain of validity, the
method promises to be easier to use than the methods mentioned above.

The basic property that characterizes Fermi liquid systems is that
the states of a noninteracting system of electrons are continuously
transformed into states of the interacting system as the interaction
strength increases from zero to its actual value.\cite{Nozieres64}
One can then study the properties of such a system by means of the
perturbation theory, regarding the interaction strength as the perturbation
parameter. Dynamics of Fermi liquid systems at low temperature and
in the linear response regime is governed by quasiparticles. However,
the question how quasiparticles propagate in a correlated system is
a non-trivial one. The answer can be extracted from the Green's function
for a particular problem if it is known. An alternative way, which
we advocate in this paper, is to analyze the excitation spectrum of
a system directly. If $E_{M}$ and $E_{M+1}$ are the ground-state
energies of an interacting ring system containing $M$ and $M+1$
electrons, respectively, the energy difference can be attributed to
the first quasiparticle energy level $\tilde{\epsilon}$ above the
Fermi energy, \begin{equation}
\tilde{\epsilon}=E_{M+1}-E_{M}.\end{equation}
The variation of the quasiparticle energy with flux threading the
ring allows us to determine the conductance of the system. The complete
proof of the formalism is given in RR and a brief overview is presented
in the next Section. Here we show how the method can be implemented
in practice. 

\begin{figure}[htbp]
\begin{center}\includegraphics[%
  width=6cm,
  keepaspectratio]{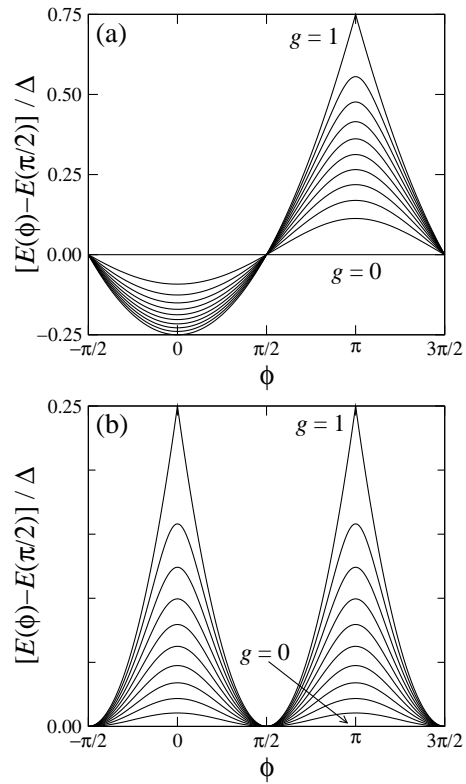}\end{center}

\caption{\label{cap:System2}The ground-state energy of an interacting system
as a function of flux $\phi$ for an even (a) and an odd (b) number
of electrons for $g$ going from 0 to 1 in steps of 0.1 and $N\to\infty$.}
\end{figure}

The key property of ring systems presented in Fig.~\ref{cap:System1}(b)
is the universality expressed in the variation of the ground-state
energy with auxiliary magnetic flux through the ring.  Here we assume
a system obeys the time reversal symmetry. The more general case is
presented in the last Section. For an even number of electrons in
the system and a large number of sites in the ring $N\to\infty$ the
ground-state energy takes a universal form\begin{equation}
E\left(\phi\right)-E\left(\frac{\pi}{2}\right)=\frac{\Delta}{\pi^{2}}\Bigl(\arccos^{2}\left(\mp\sqrt{g}\cos\phi\right)-\frac{\pi^{2}}{4}\Bigr),\label{eq:even}\end{equation}
where the average level spacing at the Fermi energy $\Delta=[N\rho\left(\epsilon_{F}\right)]^{-1}$
is determined by the density of states at the Fermi energy in an infinite
noninteracting lead $\rho\left(\epsilon_{F}\right)$ and $g=G/G_{0}$
is the dimensionless conductance. For systems with an odd number of
electrons, the ground-state energy is given with

\begin{equation}
E\left(\phi\right)-E\left(\frac{\pi}{2}\right)=\frac{\Delta}{\pi^{2}}\arcsin^{2}\left(\sqrt{g}\cos\phi\right).\label{eq:odd}\end{equation}
It should be mentioned that the ground-state energy of an interacting
ring system exactly corresponds to the expression for persistent currents
in noninteracting rings, as derived by Gogolin.~\cite{Gogolin94}
The only parameter determining the ground-state energy is the conductance
$g$ of the original system, Fig.~\ref{cap:System1}(a). In Fig.~\ref{cap:System2}
the ground-state energy as a function of the flux $\phi$ for even
and odd numbers of electrons is presented. It should be pointed out
that as successive quasiparticle levels are being occupied, the points
$\phi=0$ and $\phi=\pi$ interchange their roles (as is also the
case for noninteracting systems), and that the periodicity in even
and odd cases are $\pi$ and $\pi/2$, respectively. 

\begin{figure}[htbp]
\begin{center}\includegraphics[%
  width=6cm,
  keepaspectratio]{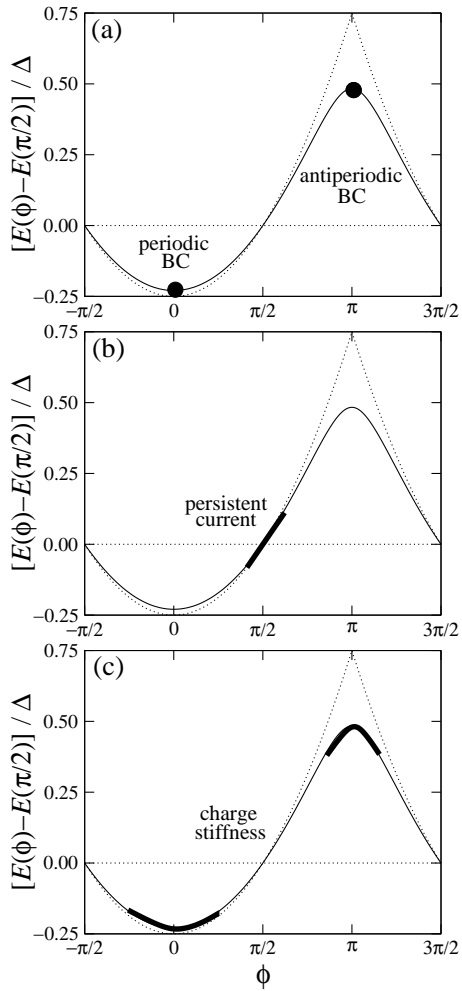}\end{center}

\caption{\label{cap:System3}The conductance can be extracted from: (a) the
two-point formula, Eq.~(\ref{eq:averageS}), (b) the persistent current
formula, Eq.~(\ref{eq:halfpiS}), or (c) the charge stiffness formula,
Eq.~(\ref{eq:stiff}).}
\end{figure}

If the ground state of the system in Fig.~\ref{cap:System1}(b) is
known, the conductance of the original open system can be extracted
from Eq.~(\ref{eq:even}) {[}or Eq.~(\ref{eq:odd}){]}. There are
several ways how to determine $g$ from Eq\emph{.~}(\ref{eq:even}).
The simplest seems to be the use of the relation
\begin{equation}
g=\sin^{2}\left(\frac{\pi}{2}\frac{E\left(\pi\right)-E\left(0\right)}{\Delta}\right),\label{eq:averageS}\end{equation}
where $E(0)$ and $E(\pi)$ are the ground-state energies of the ring
system for $\phi=0$ and $\phi=\pi$, respectively. The first advantage
of this formula is the fact that the energies can be calculated using
periodic and antiperiodic boundary conditions, respectively, and thus
the wave functions of the system can be taken real. Additional advantage
is fast convergence with $N$, as briefly discussed in the last Section.
This formula was derived in the $g\to0$ limit as $g=
\left(\frac{\pi}{2\Delta}[E\left(\pi\right)-E\left(0\right)]\right)^2$
by Favand and Mila for noninteracting
systems and applied to interacting Hubbard chains.\cite{Favand98}
More recently, a similar approach was performed in Ref.~\onlinecite{Molina02}.
In Fig.~\ref{cap:System3}(a) the use of formula Eq.~(\ref{eq:averageS})
is schematically presented.

The derivative of the ground-state energy with respect to flux gives
the persistent current in the ring $j(\phi)=\frac{e}{\hbar}\frac{\partial E}{\partial\phi}$.~\cite{Bloch65}
The second formula relates the conductance to the persistent current
at $\phi=\frac{\pi}{2}$,\begin{equation}
g=\left(\frac{\pi}{\Delta}\frac{\hbar}{e}j\left(\frac{\pi}{2}\right)\right)^{2}.\label{eq:halfpiS}\end{equation}
This relation was recently derived for noninteracting systems \cite{Sushkov01}
and successfully applied to systems with interaction.\cite{Sushkov01,Meden03} 

\begin{figure}[htbp]
\begin{center}\includegraphics[%
  width=6cm,
  keepaspectratio]{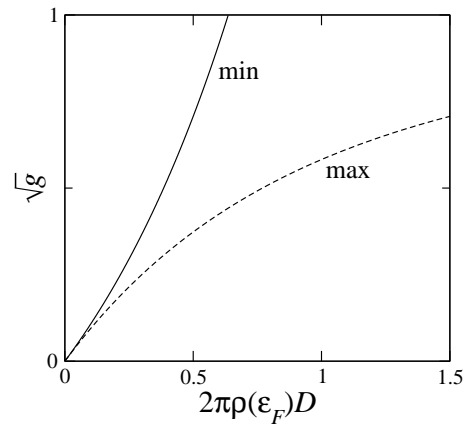}\end{center}

\caption{\label{cap:System4}Conductance vs. charge stiffness using $D$
at the energy minimum (full line) and $\tilde D$ at the maximum
(dashed line). Note the quadratic dependence $g\propto D^{2}$ for
$g\to0$.}
\end{figure}

At $T=0$ the charge stiffness is an important quantity describing
the charge transport in correlated systems.\cite{Prelovsek01} It
is defined as the second derivative of the ground-state energy of
the system with respect to the flux in the minimum of the energy vs.
flux curve, $D=\frac{N}{2}\left.\partial^{2}E/\partial\phi^{2}\right|_{E=\mathrm{min}}$.\cite{Kohn64} The sensitivity of the ground state energy to flux 
has been applied also in the context of electron localisation.\cite{edwards72}
One can also define the corresponding quantity for the energy maximum
as $\tilde{D}=-\frac{N}{2}\left.\partial^{2}E/\partial\phi^{2}\right|_{E=\mathrm{max}}$.
From Eq\emph{.~}(\ref{eq:even}) the conductance can be related to
the charge stiffness with an implicit relation,

\begin{equation}
\frac{1}{\Delta}\left.\frac{\partial^{2}E}{\partial\phi^{2}}\right|_{E=\mathrm{min,}\,\mathrm{max}}=\pm\frac{2}{\pi^{2}}\sqrt{\frac{g}{1-g}}\arccos\left(\pm\sqrt{g}\right).\label{eq:stiff}\end{equation}
Here the upper and the lower signs correspond to the second derivative
at a minimum and at a maximum of the energy vs. flux curve, respectively.
In general, this equation has to be solved numerically and the solutions
are presented in Fig.~\ref{cap:System4}. In the limit of a very
small conductance and in the vicinity of the unitary limit, analytic
formulas are available\begin{equation}
g=\left\{ \begin{array}{ll}
\left[2\pi\rho\left(\varepsilon_{F}\right)D\right]^{2}, & g\rightarrow0,\\
\left(\frac{1}{2}+\frac{3\pi}{4}\left[2\pi\rho\left(\varepsilon_{F}\right)D\right]\right)^{2}, & g\rightarrow1.\end{array}\right.\end{equation}
Note that there is a quadratic relation between the conductance and
the charge stiffness in the low conductance limit. The corresponding
formulas for the maximum of the energy vs. flux curve are\begin{equation}
g=\left\{ \begin{array}{ll}
\left[2\pi\rho\left(\varepsilon_{F}\right)\tilde{D}\right]^{2}, & g\rightarrow0,\\
\left(1-\frac{2}{\left[2\pi\rho\left(\varepsilon_{F}\right)\tilde{D}\right]^{2}}\right)^{2}, & g\rightarrow1.\end{array}\right.\end{equation}

It should be stressed that the validity of all formulas presented
in this Section is based on an assumption that the number of sites
in the ring is sufficiently large according to the condition~\cite{RR}\begin{equation}
N\gg\frac{1}{\rho\left(\varepsilon_{F}\right)}\frac{\partial\sqrt{g\left(\varepsilon_{F}\right)}}{\partial\varepsilon_{F}}.\label{cond}\end{equation}
This means that if $g\left(\varepsilon_{F}\right)$ exhibits sharp
resonances, as is the case, e.g., in chaotic systems,\cite{Jalabert92}
the calculation has to be performed on a large auxiliary ring system
and in such cases the method might be impractical compared to other
methods. On the other hand, for systems with strong interaction the
method promises to be extremely efficient already for ring systems
of a moderate size.~\cite{brr,Molina02,Meden03}

\section{Proof of the formalism}

The complete proof of the formalism is presented in RR, here we briefly
describe the main steps. The proof strongly relies on an assumption
that the ground state of the system under investigation is a Fermi
liquid.\cite{Landau56}

We start with the linear response conductance of a general interacting
system of the type shown in Fig.~\ref{cap:System1}(a). The conductance
can be calculated from the Kubo formula~\cite{Kubo57}\begin{equation}
g=\lim_{\omega\rightarrow0}\frac{i\pi}{\omega+i\delta}\Pi_{II}\left(\omega+i\delta\right),\end{equation}
where $\Pi_{II}\left(\omega+i\delta\right)$ is the retarded current-current
correlation function. For Fermi liquid systems, the current-current
correlation function can be calculated within the perturbation theory.
At $T=0$, only the bubble diagram gives a non-vanishing contribution~\cite{Oguri01}
and the conductance can be expressed in terms of the Green's function
$G_{n^{\prime}n}\left(z\right)$ of the system,

\begin{equation}
g=\left|\frac{1}{-i\pi\rho\left(\epsilon_{F}\right)}e^{-ik_{F}\left(n^{\prime}-n\right)}G_{n^{\prime}n}\left(\epsilon_{F}+i\delta\right)\right|^{2},\label{eq:FermiLB2}\end{equation}
where $n$ and $n^{\prime}$ are sites in the left and the right lead,
respectively. 

In Fermi liquid systems obeying the time-reversal symmetry,\cite{RRAB}
the imaginary part of the retarded self-energy at $T=0$ vanishes
at the Fermi energy and is quadratic for frequencies close to the
Fermi energy.\cite{Yamada86} Using the Fermi energy as the origin
of the energy scale, i.e. $\omega-\epsilon_{F}\rightarrow\omega$,
we can express this as\begin{equation}
\mathrm{Im}\Sigma_{ij}\left(\omega+i\delta\right)\propto\omega^{2}.\label{Luttinger}\end{equation}
Close to the Fermi energy, the self-energy can be expanded in powers
of $\omega$ resulting in an approximation to the Green's function,\begin{eqnarray}
 &  & \mathbf{G}^{-1}\left(\omega+i\delta\right)=\omega\mathbf{1}-\mathbf{H}^{\left(0\right)}-\bm\Sigma\left(0+i\delta\right)-\nonumber \\
 &  & \quad-\omega\left.\frac{\partial\bm\Sigma\left(\omega+i\delta\right)}{\partial\omega}\right|_{\omega=0}+\mathcal{O}\left(\omega^{2}\right).\label{eq:expansion}\end{eqnarray}
 Here $\mathbf{H}^{\left(0\right)}$ contains matrix elements of the
noninteracting part of the Hamiltonian. The Green's function for $\omega$
close to the Fermi energy can then be expressed as\begin{equation}
\mathbf{G}^{-1}\left(\omega+i\delta\right)=\mathbf{Z}^{-1/2}\tilde{\mathbf{G}}^{-1}\left(\omega+i\delta\right)\mathbf{Z}^{-1/2}+\mathcal{O}\left(\omega^{2}\right)\!,\!\!\label{eq:gfermi}\end{equation}
where we defined the quasiparticle Green's function\begin{equation}
\tilde{\mathbf{G}}^{-1}\left(\omega+i\delta\right)=\omega\mathbf{1}-\tilde{\mathbf{H}}\end{equation}
as the Green's function of a \emph{noninteracting} \emph{quasiparticle}
Hamiltonian

\begin{equation}
\tilde{\mathbf{H}}=\mathbf{Z}^{1/2}\left[\mathbf{H}^{\left(0\right)}+\bm\Sigma\left(0+i\delta\right)\right]\mathbf{Z}^{1/2},\label{quasiham}\end{equation}
 and introduced the renormalization factor matrix $\mathbf{Z}$. Matrix
elements of $\mathbf{Z}$ differ from those of an identity matrix
only if they correspond to sites of the central region. 

The reason for introducing the quasiparticle Hamiltonian is to obtain
an alternative expression for the conductance in terms of the quasiparticle
Green's function. Eq.~(\ref{eq:gfermi}) relates the values of the
true and the quasiparticle Green's function at the Fermi energy, \begin{equation}
\mathbf{G}\left(0+i\delta\right)=\mathbf{Z}^{1/2}\tilde{\mathbf{G}}\left(0+i\delta\right)\mathbf{Z}^{1/2}.\label{eq:zgtildez}\end{equation}
Specifically, $G_{n^{\prime}n}\left(0+i\delta\right)=\tilde{G}_{n^{\prime}n}\left(0+i\delta\right)$
if both $n$ and $n^{\prime}$ are sites in the leads, as a consequence
of the properties of the renormalization factor matrix $\mathbf{Z}$
discussed above. Eq.~(\ref{eq:FermiLB2}) then tells us that the
zero-temperature conductance of a Fermi liquid system is identical
to the zero-temperature conductance of a noninteracting system defined
with the quasiparticle Hamiltonian for a given value of the Fermi
energy.

These conclusions are valid if the central region is coupled to semi-infinite
leads. Here we generalize the concept of quasiparticles to a \emph{}finite
ring system with $N$ sites and $M$ electrons, threaded by a magnetic
flux $\phi$. One can define the quasiparticle Hamiltonian for such
a system,\begin{equation}
\tilde{\mathbf{H}}\left(N,\phi;M\right)=\mathbf{Z}^{1/2}\left[\mathbf{H}^{\left(0\right)}\left(N,\phi\right)+\bm\Sigma\left(0+i\delta\right)\right]\mathbf{Z}^{1/2}.\label{eq:assumptquasi}\end{equation}
 Here the self-energy and the renormalization factor matrix are determined
in the thermodynamic limit where, as we prove in RR, they are independent
of $\phi$ and correspond to those of an infinite two-lead system. 

Suppose now that we knew the exact values of the renormalized matrix
elements in the quasiparticle Hamiltonian (\ref{eq:assumptquasi}).
As this is a noninteracting Hamiltonian, we could then apply the conductance
formulas presented in the previous Section (the proof of validity
of energy formulas for noninteracting systems is given in RR and the corresponding
result for persistent currents in Ref.~\onlinecite{Gogolin94}).
Such a procedure would provide us with the exact conductance
of the original interacting system. However, to obtain the values
of the renormalized matrix elements, one needs to calculate the self-energy
of the system, which is a difficult many-body problem. In RR we study
the excitation spectrum of a finite ring system with interaction and
threaded with a magnetic flux. We show that\begin{eqnarray}
 &  & E\left(N,\phi;M+1\right)-E\left(N,\phi;M\right)=\nonumber \\
 &  & \quad=\tilde{\epsilon}\left(N,\phi;M;1\right)+\mathcal{O}\left(N^{-\frac{3}{2}}\right),\label{eq:assumpt}\end{eqnarray}
where $E\left(N,\phi;M\right)$ and $E\left(N,\phi;M+1\right)$ are
the ground-state energies of the interacting Hamiltonian for a ring
system with $N$ sites and flux $\phi$, containing $M$ and $M+1$
electrons, respectively, and $\tilde{\epsilon}\left(N,\phi;M;1\right)$
is the energy of the first single-electron level above the Fermi energy
of the finite ring quasiparticle Hamiltonian (\ref{eq:assumptquasi}).
The error in Eq.~(\ref{eq:assumpt}) is small enough that the proof
of the ground-state energy formulas for noninteracting systems, which
involves only the properties of a set of neighboring single-electron
energy levels, remains valid also for interacting Fermi liquid systems,
provided a system is a Fermi liquid for all values of the Fermi energy
below its actual value.

\section{Examples }

\subsection{Noninteracting system}

\begin{figure}[htbp]
\begin{center}\includegraphics[%
  width=6cm,
  keepaspectratio]{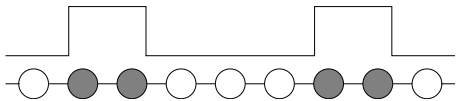}\end{center}

\caption{\label{cap:testnisys}A double barrier noninteracting system. The
height of the barriers is $0.5t$, where $t$ is the hopping matrix
element between neighboring sites.}
\end{figure}

\begin{figure}[htbp]
\begin{center}\includegraphics[%
  width=6cm,
  keepaspectratio]{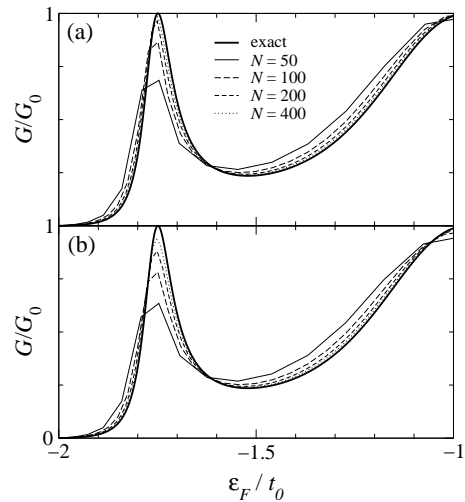}\end{center}

\caption{\label{cap:testni}(a) The conductance through the system in Fig.~\ref{cap:testnisys}
calculated from the two-point formula, Eq.~(\ref{eq:averageS}),
and (b) from the persistent current formula, Eq.~(\ref{eq:halfpiS}).
Note the different convergence behavior of the two formulas.}
\end{figure}

In this Section we discuss the convergence properties of the conductance
formulas. As the first example we take a double-barrier potential
scattering problem presented in Fig.~\ref{cap:testnisys}. Results
of various formulas for different number of sites in the ring are
presented in Fig.~\ref{cap:testni}. The exact zero-temperature conductance
for this system exhibits a sharp resonance peak superimposed on a
smooth background conductance. We notice immediately that as the number
of sites in the ring increases, the convergence is generally faster
in the region where the conductance is smooth than in the resonance
region, which is consistent with the condition Eq.~(\ref{cond}).
Comparing the results obtained employing the two-point formula Eq.~(\ref{eq:averageS})
and the persistent current formula Eq.~(\ref{eq:halfpiS}) we observe
that the convergence is better for the two-point formula expressing
the conductance in terms of the difference of the energies at $\phi=0$
and $\phi=\pi$. From the computational point of view there is an
additional advantage of the two-point formula. In this case, all the
matrix elements can be made real if one chooses such a vector potential
that only one hopping matrix element if modified by the flux as then
the additional phase factor is $e^{\pm i\pi}=-1$. 

\subsection{Anderson impurity model}

\begin{figure}[htbp]
\begin{center}\includegraphics[%
  width=5cm,
  keepaspectratio]{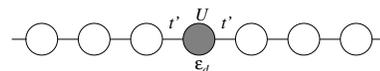}\end{center}

\caption{\label{cap:andsys}The Anderson impurity model realized as a quantum
dot coupled to two leads. The dot is described with the energy level
$\epsilon_{d}$ and the Coulomb energy of a doubly occupied level
$U$. $t^{\prime}$ is the hopping between the dot and leads.}
\end{figure}

\begin{figure}[htbp]
\begin{center}\includegraphics[%
  width=6cm,
  keepaspectratio]{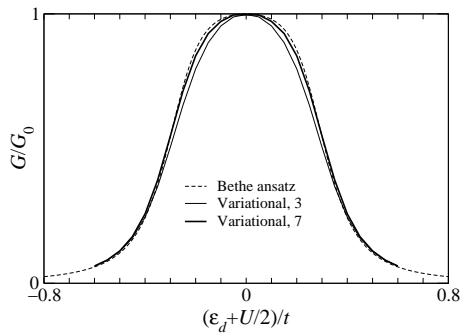}\end{center}

\caption{\label{cap:andvar}The zero-temperature conductance calculated from
ground-state energy vs. magnetic flux in a finite ring system using
the variational method described in RR with 3 and 7 basis functions.
For comparison, the exact Bethe ansatz result is presented with a
dashed line. The system is shown in Fig.~\ref{cap:andsys}, with
$U=0.64t$ and $t^{\prime}=0.2t$.}
\end{figure}

As a nontrivial example of the use of the formalism we calculate the
zero-temperature conductance of a single impurity Anderson model realized
as a quantum dot attached to leads as shown in Fig.~\ref{cap:andsys}.
In Fig.~\ref{cap:andvar} the results are compared to exact conductance
of the Bethe ansatz approach.\cite{Wiegman82a} To calculate the conductance,
Eq.~(\ref{eq:averageS}) was used, with the ground-state energies
at $\phi=0$ and $\phi=\pi$ obtained using a variational method described
in RR. For each position of the $\epsilon_{d}$ level relative to
the Fermi energy, we increased the number of sites in the ring until
the conductance converged. The number of sites needed to achieve the
convergence was the lowest in the empty orbital regime and the highest
(about 1000 for the system shown in Fig.~\ref{cap:andvar}) in the
Kondo regime. This is a consequence of Eq.~(\ref{cond}) as a narrow
resonance related to the Kondo resonance appears in the transmission
probability of the quasiparticle Hamiltonian (\ref{quasiham}) in
the Kondo regime. 

\subsection{Interacting Aharonov Bohm rings}

\begin{figure}[htbp]
\begin{center}\includegraphics[%
  bb=200bp 0 620bp 652bp,
  clip,
  width=7cm,
  keepaspectratio]{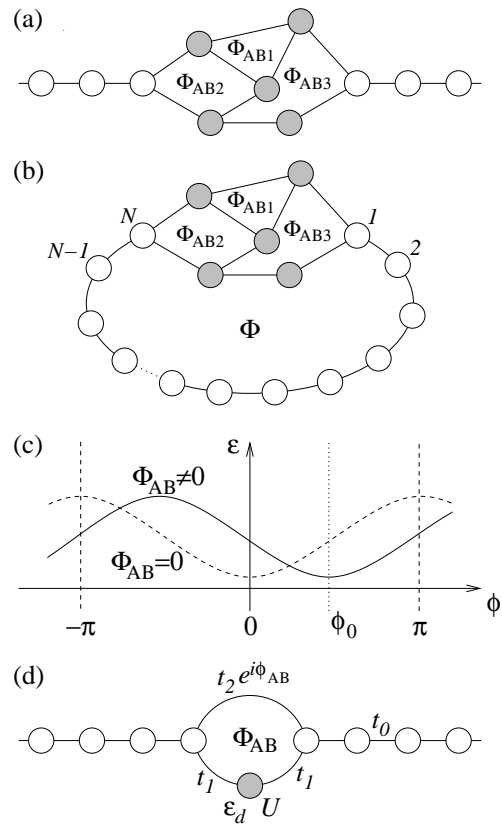}\end{center}

\caption{\label{cap:ab1}(a) Interacting mesoscopic region (gray-shaded sites),
threaded by magnetic flux and coupled to noninteracting leads. (b)
Auxiliary ring system. (c) Behavior of energy levels as the flux threading
the ring is varied. (d) Example system with a quantum dot embedded
in an AB ring.}
\end{figure}
One can generalize the conductance formulas to systems which exhibit
time reversal asymmetry, such as is e.g. an Aharonov-Bohm (AB) type
of system presented in Fig.~\ref{cap:ab1}(a).\cite{RRAB} If there
is no AB flux threading the mesoscopic region, the time reversal symmetry
is restored and the energy is an even function of $\phi$. In the
general case, the energy extremum is shifted to a non-trivial point
$\phi_{0}\left(\varepsilon_{F}\right)$, as illustrated in Fig.~\ref{cap:ab1}(c).
The ground-state energy is then generalized to $E(\phi)=\pi^{-2}\Delta\arccos^{2}\left(\mp\sqrt{g}\cos\left[\phi-\phi_{0}\left(\varepsilon_{F}\right)\right]\right)+\mathrm{const.}$
for an even number of electrons in a system and to $E(\phi)=\pi^{-2}\Delta\arcsin^{2}\left(\sqrt{g}\cos\left[\phi-\phi_{0}\left(\varepsilon_{F}\right)\right]\right)+\mathrm{const.}$
for an odd number of electrons in a system. From the former expression,
the transmission probability can be extracted, and the conductance
is given by~\cite{RRAB}\begin{equation}
g=\sin^{2}\left(\frac{\pi}{2}\frac{E\left(\phi_{0}+\pi\right)-E\left(\phi_{0}\right)}{\Delta}\right),\label{eq:gsin2}\end{equation}
where $\phi_{0}\equiv\phi_{0}\left(\epsilon_{F}\right)$ is determined
by the position of the minimum (or maximum) in the energy vs. flux
curve. The conductance can also be calculated from the more convenient
four-point formula~\cite{3point}

\begin{equation}
g=\sin^{2}\!\!\left(\!\frac{\pi}{2}\frac{E\!\left(\pi\right)\!-\! E\!\left(0\right)}{\Delta}\!\right)\!\!+\!\sin^{2}\!\!\left(\!\frac{\pi}{2}\frac{E\!\left(\pi/2\right)\!-\! E\!\left(\!-\pi/2\right)}{\Delta}\!\right),\label{eq:gsin3}\end{equation}
and $\phi_{0}$ is determined with the expression\begin{equation}
\phi_{0}=-\arctan\frac{\sin\left(\frac{\pi}{2}\frac{E\left(\pi/2\right)-E\left(-\pi/2\right)}{\Delta}\right)}{\sin\left(\frac{\pi}{2}\frac{E\left(\pi\right)-E\left(0\right)}{\Delta}\right)}.\label{eq:phi0}\end{equation}
If no AB flux is present in the mesoscopic region, we recover the
two-point formula Eq.~(\ref{eq:averageS}) since $E\left(\pi/2\right)=E\left(-\pi/2\right)$
in this case. 

If the time reversal symmetry is broken due to AB flux, Eq.~(\ref{Luttinger})
is not valid and the proof of the previous Section has to be reconsidered.
Repeating the steps as presented in detail in RR, the proof is restored
and basically unchanged if the self energy obeys the relation\begin{equation}
\frac{1}{2i}\left[\Sigma_{ij}\left(\omega+i\delta\right)-\Sigma_{ij}\left(\omega-i\delta\right)\right]\propto\omega^{2}.\label{Luttinger2}\end{equation}
It follows that the linear response conductance of an interacting
AB system at zero temperature is given by the four-point formula Eq.~(\ref{eq:gsin3}).
The condition Eq.~(\ref{Luttinger2}) is fulfilled if the system
is a Fermi liquid. 

\begin{figure}[htbp]
\begin{center}\includegraphics[%
  width=6cm,
  keepaspectratio]{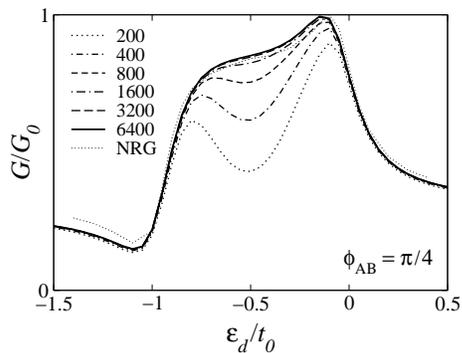}\end{center}

\caption{\label{cap:ab2}Zero-temperature linear response conductance of the
system in Fig.~\ref{cap:ab1}(d) as a function of level position
$\varepsilon_{d}$ for various number of ring sites $N$ and $\phi_{AB}=\pi/4$.
The dotted line is the NRG result from Ref.~\onlinecite{Hofstetter01}.
Parameters: $t_{1}=0.177t_{0}$ ($\Gamma=0.125t_{0}$), $t_{2}=0.298t_{0}$,
$U=t_{0}=8\Gamma$. }
\end{figure}

In order to demonstrate the practical value of the method, we quantitatively
analyze the conductance through an Aharonov-Bohm ring with a quantum
dot placed in one of the arms\cite{Bulka01,Hofstetter01} as presented
in Fig.~\ref{cap:ab1}(d). In Fig.~\ref{cap:ab2} a convergence
test of the method is shown. The convergence with $N$ is fast in
the empty orbital regime and becomes progressively slower as $\epsilon_{d}$
shifts toward the Kondo regime, as was also the case for a quantum
dot attached to leads. The converged conductance curve is in excellent
agreement with the numerical renormalization group result of Ref.~\onlinecite{Hofstetter01}.

\section{Summary}

We have demonstrated how the zero-temperature conductance of a sample
with electron-electron interaction, attached to noninteracting leads
can be determined. The method is extremely simple and is based on
several formulas relating the conductance to the ground-state energy
of an auxiliary ring system. The conductance is determined from the
ground-state energy of an interacting system, while in more traditional
approaches, one needs to know the Green's function of the system.
The advantage of the present method comes from the fact that the ground-state
energy is often relatively simple to obtain by various numerical approaches,
including variational methods. Let us summarize the key points of
the method:

(1) The {}``open'' problem of the conductance through a sample coupled
to semi-infinite leads is mapped on to a ''closed'' problem, namely
a ring threaded by a magnetic flux and containing the same correlated
electron region. 

(2) For the case of a Fermi liquid interacting system, even with broken
time-reversal symmetry, it is shown that the zero-temperature conductance
can be deduced from the variation of the ground state energy with
the flux in a large, but finite ring system. 

(3) In order to prove this, the concept of Fermi liquid quasiparticles
is extended to finite, but large systems. The conductance formulas
give the conductance of a system of noninteracting quasiparticles,
which is equal to the conductance of the original interacting system.

(4) The results of our method are compared to results of other approaches
for problems such as the transport through a quantum dot containing
interacting electrons. The comparison shows an excellent quantitative
agreement with exact Bethe ansatz results. We have demonstrated the
usefulness of the formula also by applying it to a prototype system
exhibiting Kondo-Fano behavior. Results based on the four-point formula
confirm the results of the numerical renormalization group method.


\begin{thebibliography}{10}
\bibitem{Delft01}J. von Delft and D.~C. Ralph, Phys. Rep. \textbf{345}, 61 (2001).
\bibitem{Kouwenhoven97}L.~P. Kouwenhoven \textit{et~al.}, in \emph{Mesoscopic Electron
Transport}, edited by L.~L. Sohn, L.~P. Kouwenhoven, and G. Sch\"{o}n
(Kluwer Academic, New York, 1997).
\bibitem{Agrait02}N. Agra\"{\i}t, A.~L. Yeyati, and J.~M. van Ruitenbeek, Phys. Rep.
\textbf{377}, 81 (2003).
\bibitem{Smit02}R.~H.~M. Smit \textit{et~al.}, Nature \textbf{419}, 906 (2002).
\bibitem{nature}W. Liang \emph{et~al.}, Nature \textbf{417}, 725 (2002).
\bibitem{Landauer57}R. Landauer, IBM J. Res. Dev. \textbf{1}, 233 (1957); Philos. Mag.
\textbf{21}, 863 (1970); M. B\"{u}ttiker, Phys. Rev. Lett. \textbf{57},
1761 (1986).
\bibitem{Oguri97a}A. Oguri, Phys. Rev. B \textbf{56}, 13422 (1997).
\bibitem{Oguri01}A. Oguri, J. Phys. Soc. Jpn. \textbf{70}, 2666 (2001).
\bibitem{Meir92}Y. Meir and N.~S. Wingreen, Phys. Rev. Lett. \textbf{68}, 2512 (1992);
H.~M. Pastawski, Phys. Rev. B \textbf{46}, 4053 (1992).
\bibitem{abinitio}N. D. Lang and Ph. Avouris, Phys. Rev. Lett. \textbf{84}, 358 (2000);
J. Heurich \emph{et~al., ibid}. \textbf{8}8, 256803 (2002).
\bibitem{RR}T. Rejec and A. Ram\v{s}ak, Phys. Rev. B \textbf{68}, 035342 (2003).
\bibitem{Nozieres64}P. Nozi\` eres, \emph{Theory of interacting Fermi systems} (W. A. Bejnamin,
Inc., New York, 1964).
\bibitem{RRAB}T. Rejec and A. Ram\v{s}ak, Phys. Rev. B \textbf{68}, 033306 (2003).
\bibitem{Gogolin94}A.~O. Gogolin and N.~V. Prokof'ev, Phys. Rev. B \textbf{50}, 4921
(1994).
\bibitem{Favand98}J. Favand and F. Mila, Eur. Phys. J. B \textbf{2}, 293 (1998).
\bibitem{Molina02}R.~A. Molina \textit{et~al.},  Phys. Rev. B \textbf{67}, 235306
(2003).
\bibitem{Bloch65}F. Bloch, Phys. Rev. \textbf{137}, A787 (1965); M. B\"{u}ttiker, Y. Imry,
and R. Landauer, Phys. Lett. \textbf{96A}, 365 (1983).
\bibitem{Sushkov01}O.~P. Sushkov, Phys. Rev. B \textbf{64}, 155319 (2001). 
\bibitem{Meden03}V. Meden and U. Schollw\"{o}ck, Phys. Rev. B \textbf{67}, 193303 (2003).
\bibitem{Prelovsek01}P. Prelov\v{s}ek and X. Zotos, in \emph{Lectures on the Physics of Highly
Correlated Electron Systems VI}, edited by F. Mancini (American Institute
of Physics, New York, 2002).
\bibitem{Kohn64}W. Kohn, Phys. Rev. \textbf{133}, A171 (1964).
\bibitem{edwards72} J.T. Edwards and D.J. Thouless, J. Phys. C{\bf 5}, 807 (1972).
\bibitem{Jalabert92}R.~A. Jalabert, A.~D. Stone, and Y. Alhassid, Phys. Rev. Lett. \textbf{68},
3468 (1992); J.~A. Verges, E. Cuevas, M. Ortu\~{n}o, and E. Louis,
Phys. Rev. B \textbf{58}, R10143 (1998).
\bibitem{brr}J. Bon\v{c}a, A. Ram\v{s}ak, and T. Rejec, unpublished.
\bibitem{Landau56}L.~D. Landau, Pis'ma Zh. Eksp. Teor. Fiz. \textbf{3}, 920 (1956);
\textbf{5}, 101 (1957); P. Nozi\' eres, J. Low Temp. Phys. \textbf{17},
31 (1974).
\bibitem{Kubo57}R. Kubo, J. Phys. Soc. Jpn. \textbf{12}, 570 (1957).
\bibitem{Yamada86}K. Yamada and K. Yosida, Prog. Theor. Phys. \textbf{76}, 621 (1986);
A. Oguri, J. Phys. Soc. Jpn. \textbf{66}, 1427 (1997).
\bibitem{Fisher81}D.~S. Fisher and P.~A. Lee, Phys. Rev. B \textbf{23}, 6851 (1981).
\bibitem{Wiegman82a}P.~B. Wiegman and A.~M. Tsvelick, Pis'ma Zh. Eksp. Teor. Fiz. \textbf{35},
100 (1982); J. Phys. C \textbf{16}, 2281 (1983).
\bibitem{Bulka01}B.~R. Bu\l ka and P. Stefanski, Phys. Rev. Lett. \textbf{86}, 5128
(2001).
\bibitem{Hofstetter01}W. Hofstetter, J. K\"{o}nig and H. Schoeller, Phys. Rev. Lett. \textbf{87},
156803 (2001).
\bibitem{3point}It is also possible to calculate the conductance from the ground-state
energies at three distinct values of flux $\phi$ by solving equations
numerically.\end{thebibliography}
\end{document}